\documentclass[preprint,12pt]{elsarticle}

\usepackage{amssymb}
\usepackage{graphicx}
\usepackage{epstopdf}

\journal{Astroparticle Physics}

\begin{document}

\begin{frontmatter}


\title{\tnoteref{label1}}
\author{Ryan Price\corref{cor1}\fnref{label2}}
 \ead{ryan.g.priceATutah.edu}
\author{Stephane Vincent\fnref{label2}}
\author{Stephan LeBohec\fnref{label2}}
 \address{Department of Physics and Astronomy, University of Utah, Salt-Lake-City, UT 84112-0830, USA\fnref{label2}}

\title{Haar wavelets as a tool for the statistical characterization of variability}



\begin{abstract}
In the field of gamma-ray astronomy, irregular and noisy datasets make difficult the characterization of light-curve features in terms of statistical significance while properly accounting for trial factors associated with the search for variability at different times and over different timescales. In order to address these difficulties, we propose a method based on the Haar wavelet decomposition of the data. It allows statistical characterization of possible variability, embedded in a white noise background, in terms of a confidence level. The method is applied to artificially generated data for characterization as well as to the the very high energy M87 light curve recorded with VERITAS in 2008 which serves here as a realistic application example.   

\end{abstract}


\end{frontmatter}


\section{Introduction}
\label{intro}
In high energy astrophysics  and possibly in other fields of research, problems associated with the  analysis of data with larger statistical errors (with rarely more than a few standard deviations per data point) lead to the development of very systematic and careful practices when determining the detection of a new source.  As an example, for a discovery to be claimed, because of always possible systematic effects, unrealistically high statistical significance thresholds  of $\sim5$ standard deviations above the background \cite{LiMa83} are typically self-imposed \cite{weekes, aharonian}. Less systematic and also less stringent practices are however often applied when the analysis investigates higher order features such as time variability or energy spectrum breaks (e.g. \cite{m87veritas, m87hess, m87magic}). We are here investigating the potential of using a wavelet analysis in these situations. 

In this paper, the data will be considered as a function of time, but the analysis method can easily be used on data that is a function of any variable.  (In fact, this method could be generalized to data recorded as a function of more than one parameter such as with significance maps, but this is not to be investigated here).  The variability may concentrate on individual data points or over broad intervals.  In either case, the aim is to establish a Confidence Level (CL) with which variability can be reported.  An approach sometimes taken for this is to determine if a data model, such as a constant value, fits the data with a satisfying $\chi^2$ (e.g. \cite{acciari2008, acciari2010a, acciari2010b}).  If it does, then variability is generally not further investigated with the same data.  However, this can be misleading, and does not make full use of the data as these statistical characterizations do not depend on the order in which the data points occurred.  

Wavelet analysis should then be a better tool to use in this type of situations. The approach consists of describing the dataset by a linear combination of wavelets. A wavelet family is a complete and orthonormal functional basis whose members reach their largest magnitudes over compact domains in both their time and frequency representations\cite{numrec}. Wavelet families are generally organized in subsets corresponding to different scales of variability. Within each subset, each wavelet corresponds to a different position in the dataset. When errors are associated with the original data, they can be propagated through the wavelet coefficients calculation. The ratio between the wavelet coefficient value and its error can be regarded as the statistical significance of the contribution of that wavelet to the original data. It is related to the CL with which variability at the corresponding position and scale in the data sequence is observed.  Generally, as is the case with the commonly used Daubechies wavelets\cite{daub}, different wavelets within one timescale overlap and their coefficients are not statistically independent.  For this reason, this paper presents an analysis, which utilizes the most compact of the Daubechies wavelet familly known as D2 and usually referred to as the Haar wavelets \cite{haar1910}, which do not suffer from this inconvenience and which have recently seen a regain of interest in the domains of signal processing and optimal control of linear time varying systems\cite{optlinear}. 

In section \ref{haar} we describe the Haar wavelet basis and some of its properties. Section \ref{examples} presents a few examples and applications based on simulated data to characterize the sensitivity of the wavelet analysis relative to a standard $\chi^2$ test.  Finally, section \ref{analysis} presents a few caveats concerning the implementation and interpretation of a Haar wavelet analysis when applied to real data, and includes an application to real data from the VERITAS gamma ray observatory.

\section{Analysis with the Haar wavelets}
\label{haar}
\begin{figure}
\begin{center}
\rotatebox{0}{\includegraphics[scale=0.5]{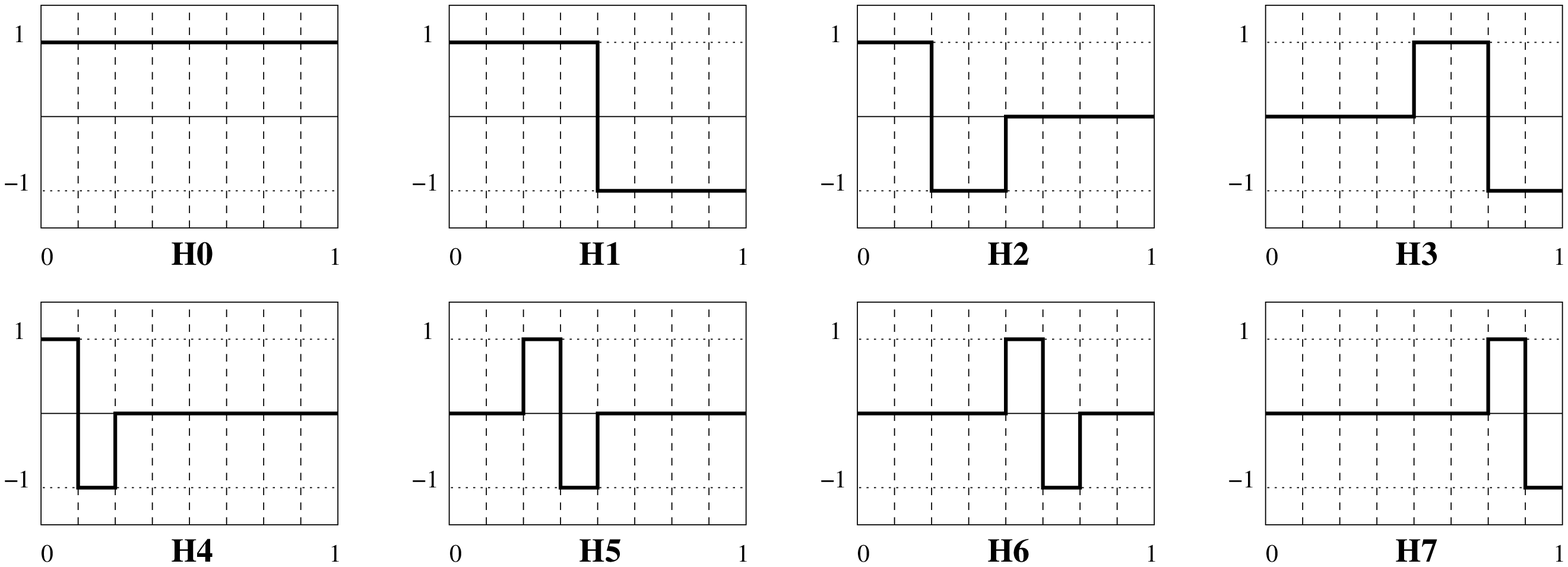}}
\end{center}
\caption{\label{haarwavelets} The first eight Haar wavelet functions. }
\end{figure}

Consider a dataset consisting of data points $s_i$ with $i=0,1,2, \cdots, N$ where the number of data points is $N=2^p$ with $p$ an integer.  The Haar wavelet coefficients $\{c_i\}$ with $i=0,1,2, \cdots, N$ can be seen as the results of functional inner products of the data sequence with the Haar functions $\{H_i\}$,  the first few of which are represented in Figure \ref{haarwavelets}. The calculation of the Haar wavelet coefficients is fairly straightforward.  The first one, $c_0$ is the average of all the data points (See $H_0$ on Figure \ref{haarwavelets}). The next one, $c_1$ is the difference between the averages over the first and second halves of the data (See $H_1$). Then, $c_2$ is the difference between the averages over the first and second quarters of the dataset (See $H_2$) while $c_3$  is the difference between the averages over the third and fourth quarters (See $H_3$). Then we go to a smaller scale with $c_4$, which is the difference between the averages of the first and second eighth of the dataset (See $H_4$). This goes on until differences between individual data points are all recorded. Clearly, there are $N/2$ coefficients measuring the differences between consecutive data points, there are $N/4$ coefficients measuring the differences between consecutive averages of groups of two data points, and there are $N/2^m$ coefficients measuring the differences between consecutive averages over groups of $2^{m-1}$ data points.

The coefficients $\{c_i\}$ are linear combinations of the data points $\{s_i\}$ to which measurement errors $\{\delta s_i\}$ may be associated. Under the assumption that the errors are gaussian and independent, it is straight forward to propagate the errors through the coefficients calculation and obtain errors $\{\delta c_i\}$ for each coefficient. The ratio $| c_i/\delta c_i |$ can be seen as the statistical significance with which variability is observed at the corresponding scale and time in the dataset. Since the Haar coefficients within a given timescale do not depend on any common data, they are statistically independent so they can be used in a $\chi^2$ analysis with $\chi^2=\sum (c_i/\delta c_i)^2$.  In the absence of variability, the expected coefficient values are zero. With the number of coefficients characterizing one specific timescale as the number of degrees of freedom, the complement of the $\chi^2$ probability provides the CL with which variability over that  timescale may be reported. 

The statistical independence of the Haar wavelet coefficients within each timescale is one of their advantages. Another advantage is  their computational and conceptual simplicity.  The Haar wavelets  provide coefficients, which can be directly understood and described in simple terms of the up and down variations from one region to the next.  However, the sensitivity to a variability feature with a given timescale  may depend on the precise timing of that feature. For example, if two consecutive data points have a higher signal than the rest of the data, the corresponding variability should ideally appear in the two data point scale. If the first of the two data points has an odd  index (with the first data point indexed as 1) the variability will indeed appear in one coefficient of the two data point scale. However, if the first of the two data points has an even index, the variability will show up in two coefficients of the one data point scale. This corresponds to an effective blurring of the specific timescale to be associated to the different wavelet scales. Section \ref{examples} shows simulated data and the corresponding Haar wavelet transforms.  

\section{Application to simulated data}
\label{examples}
In order to illustrate the Haar wavelet analysis method, dimensionless datasets of 64 points in arbitrary units (a.u.) were randomly generated. It is assumed the data points are recorded at equal time intervals, which we will use as a time unit. The time between two consecutive data points is used as the time unit. Two examples are presented here. In both cases a mean baseline at a level of 0.3\,a.u. is superimposed with a Gauss function shaped modulation. The data points are drawn randomly according to a Gauss deviate of standard deviation equal to unity and the errors  are all taken equal to 1.0\,a.u.. 

\subsection{Slow variability example}
\begin{figure}
\begin{center}
\rotatebox{0}{\includegraphics[scale=0.32]{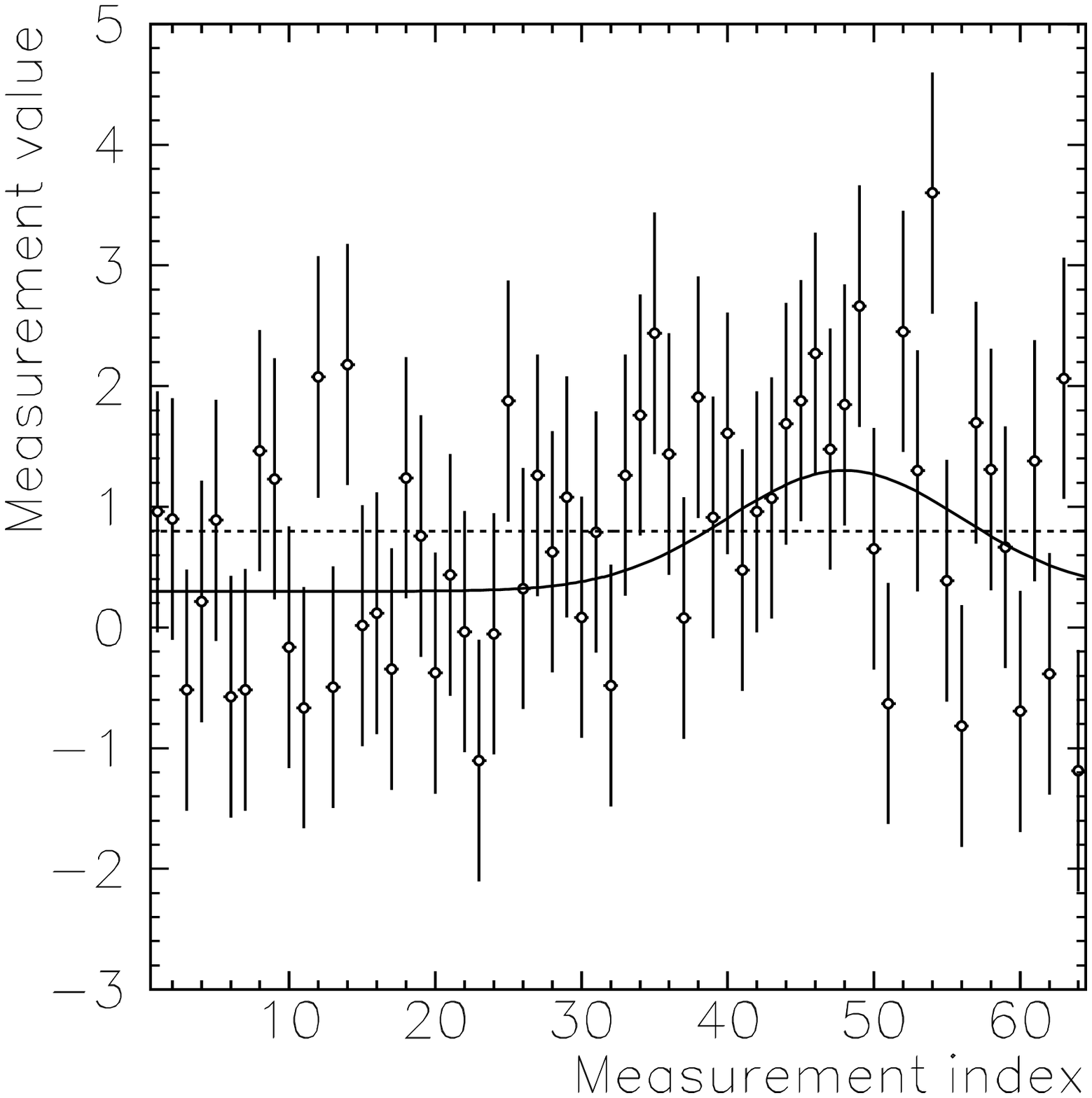}}
\rotatebox{0}{\includegraphics[scale=0.32]{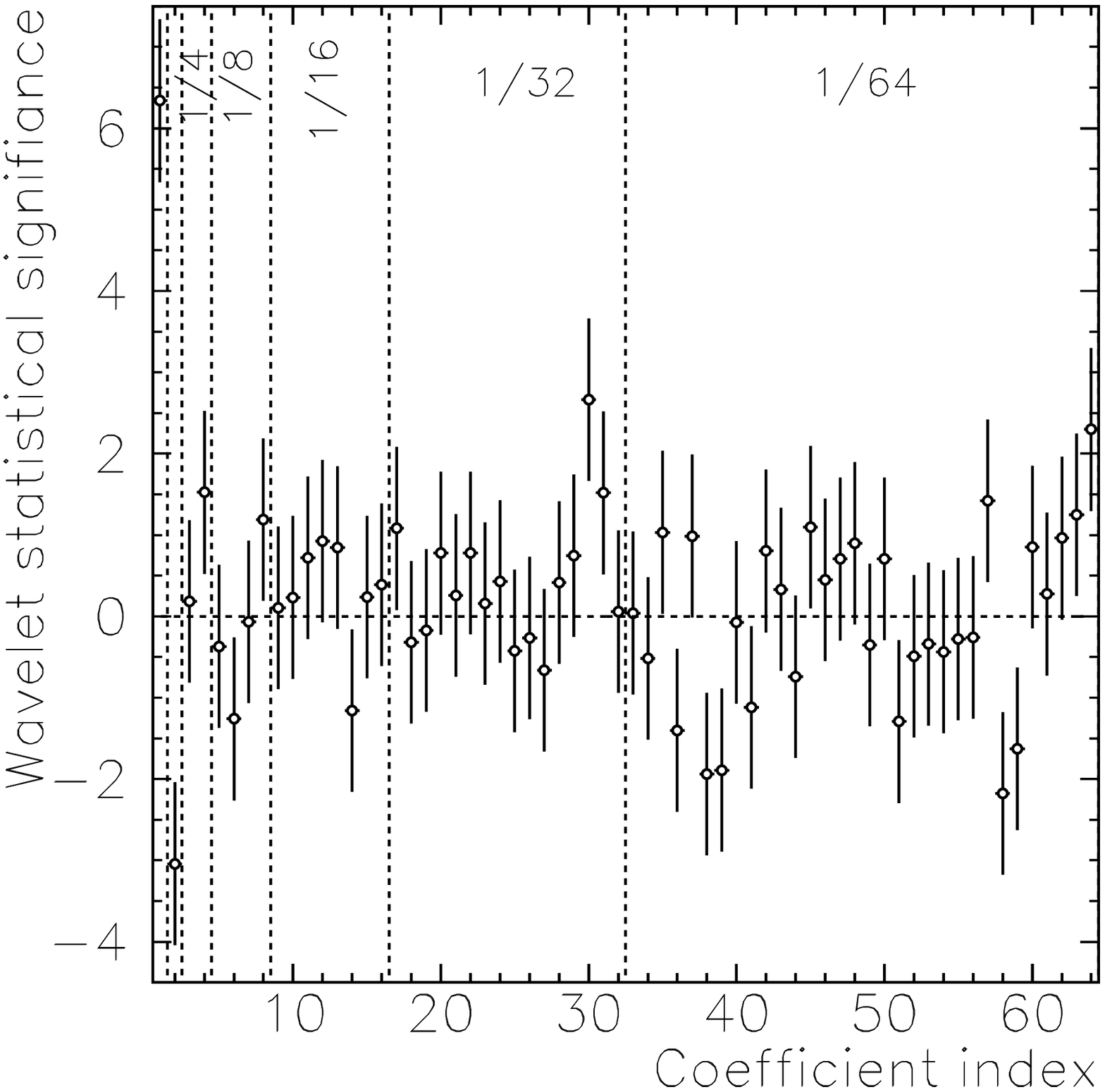}}
\end{center}
\caption{\label{figure:coef1} Left: Randomly generated data with a Gauss modulation of amplitude 1.0 centered on data point 48 with a standard deviation of 8 time units. The simulated modulation is indicated by the solid curve and the data average by the dashed line.  Right: The significance of the Haar wavelet coefficients (coefficients divided by their statistical error) arranged in groups (separated by vertical dashed lines) the longest scale on the left to the shortest on the right. The second coefficient being different from zero and negative indicates a signal increase from the first half to the second half of the dataset.}
\end{figure}
The data presented on the left panel of Figure \ref{figure:coef1} corresponds to a Gauss modulation of amplitude 1.0\,a.u. centered on data point 48 with a standard deviation of 8 time units, which is indicated by the solid line. The visual inspection of the data suggests the signal increases in the second half of the dataset.  However, the data deviates from the average (represented by the horizontal dashed line) with a very acceptable reduced $\chi^2$ of 1.109 for 63 degrees of freedom corresponding to a $\chi^2$ probability of more than 0.25 or a CL for variability of less than 75\%, which is insufficient to report any variability. 

The Haar wavelet coefficients statistical significances are shown in the right panel of Figure \ref{figure:coef1}. It is the coefficient value divided by its error that is represented and therefore, all the error bars on the graph have a unit amplitude. The first coefficient differs from zero with a statistical significance of 6.3, indicating the average signal is different from zero. The second coefficient is different from zero with a statistical significance of 3.04, indicating an increase of the signal from the first half to the second half of the data points (since the coefficient is negative, see $H_1$ on Figure \ref{haarwavelets}) with a $\chi^2$ probability or CL of more than 99.7\%. This is to be compared to the CL of 75\% obtained from the simple $\chi^2$ test applied on the original data, and indicates a greater sensitivity of the Haar wavelet analysis. The second half of the groups of coefficients in the $1\over16$ and $1\over32$ scales (four and two time units)  display some deviation from zero but not at a sufficiently significant level to reveal actual variability at these timescales. The global $\chi^2$ probabilities under the null hypothesis are 69\% and 46\% respectively for these two timescales.  Choosing a timescale, for example the one with the largest $\chi^2$ probability for variability, must be associated with a trial factor penalty. However, the wavelet coefficients in different timescales are not statistically independent. Further more, a same variability feature is likely to affect the coefficients in more than one timescale. As a consequence,  the CL obtained for each and all timescales should be reported with their timescale of relevance.

\subsection{Fast variability example}
\begin{figure}
\begin{center}
\rotatebox{0}{\includegraphics[scale=0.32]{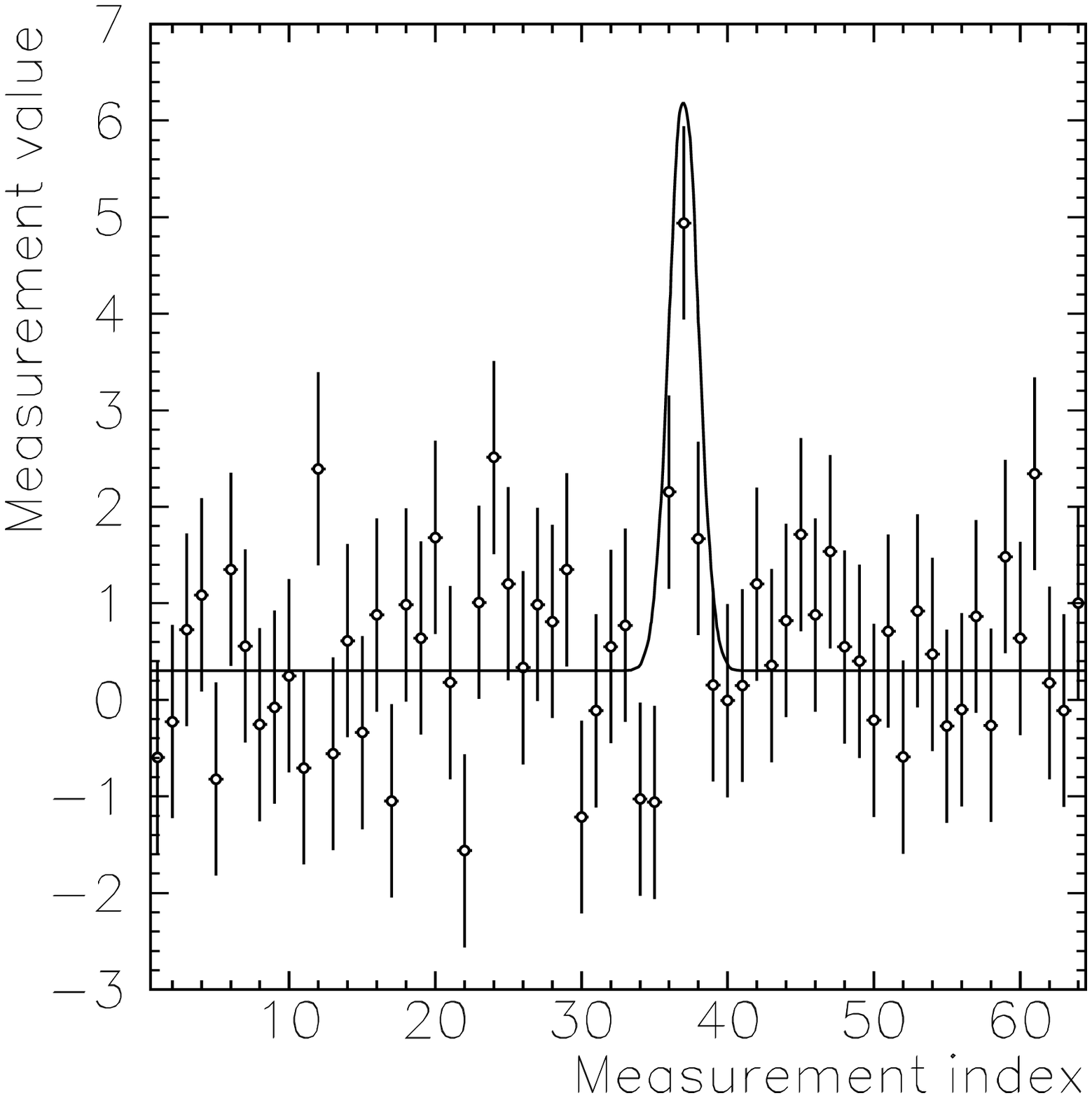}}
\rotatebox{0}{\includegraphics[scale=0.32]{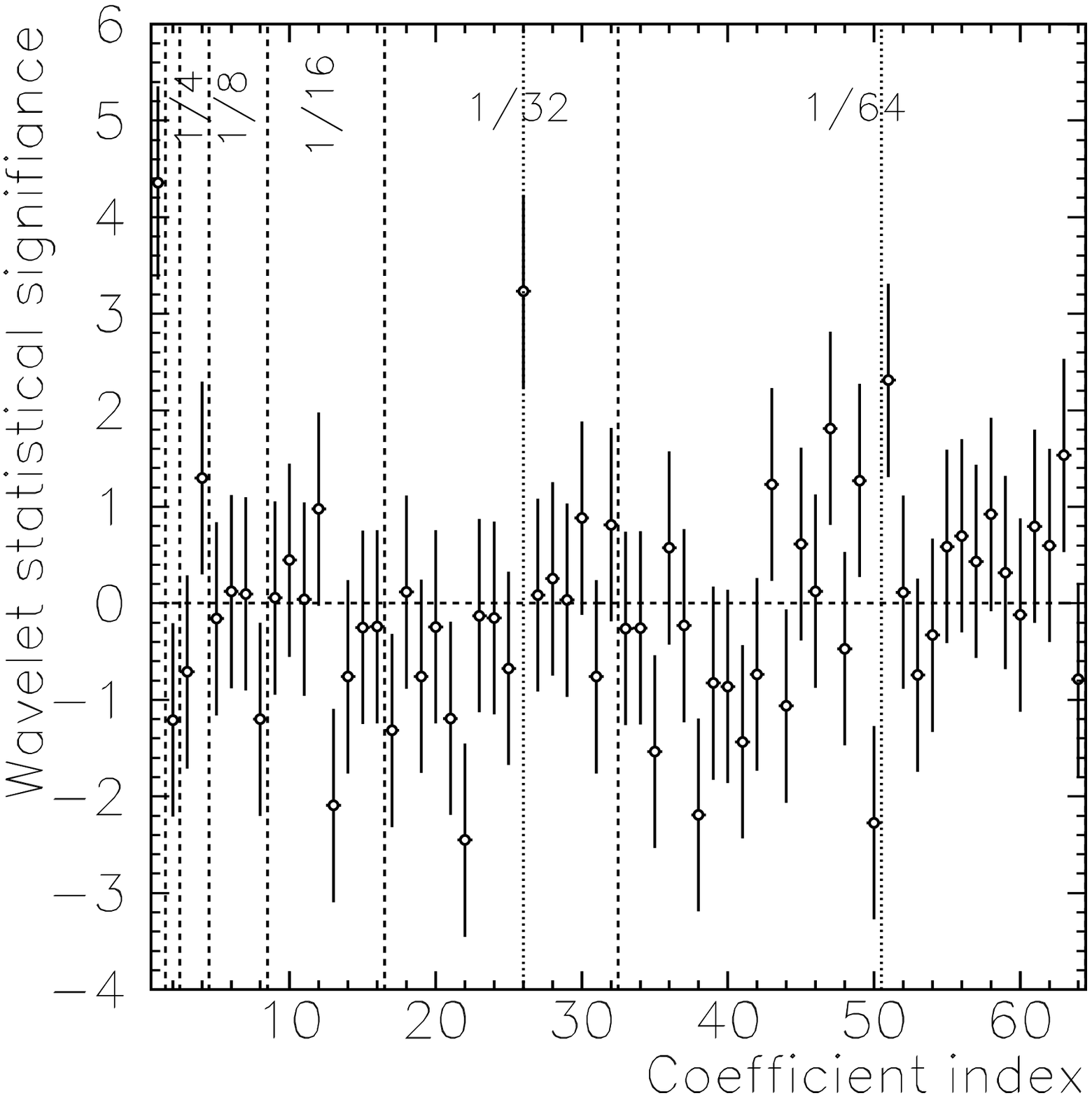}}
\end{center}
\caption{\label{figure:coef4} Same as Figure \ref{figure:coef1} for a modulation of amplitude 6.0, centered on point 37 and a standard half width corresponding to 1 time unit. Although the pulse appears in the data as well as in the Haar wavelet coefficients statistical significance (see the vertical dotted lines), the CL associated with the modulation is not sufficient for any variability to be established. }
\end{figure}

Figure \ref{figure:coef4} presents the result of a similar simulation with a Gauss modulation of amplitude 6.0\,a.u., centered on data point 37 with a standard deviation corresponding to one time unit. Visual inspection draws attention to the simulated variability with one point at almost five standard deviations from the baseline and its neighbors at more than one standard deviation. As long as external information (such as simultaneous observations in other energy bands) is not available, the CL in the identification of this feature must fully account for the implicit trial factor associated with the selection of a specific position and scale in the data  among all the possibilities. The $\chi^2$ test for variability gives a reduced $\chi^2$ of 1.13 for 63 degrees of freedom, corresponding to a $\chi^2$ probability of more than 0.22 or a CL of less than 78\%. The right panel of Figure \ref{figure:coef4} presents the statistical significances of the Haar wavelet coefficients. The simulated modulation appears in one coefficient standing out at more than three standard deviations in the $1\over 32$  scale (two time units). It may also appear in the $1\over 64$ scale (one time unit) with a couple of coefficients at a little more than two standard deviation from zero. These could be interesting as they start to resolve the structure of the pulse. However, CL for these scales are of 88\% and 75\%, insufficient for variability to be reported. The Haar wavelet analysis does not provide better result than a $\chi^2$ test for the shortest variability timescales.

\subsection{Statistical comparison with the direct $\chi^2$ test for variability}

\begin{figure}
\begin{center}
\rotatebox{0}{\includegraphics[scale=0.32]{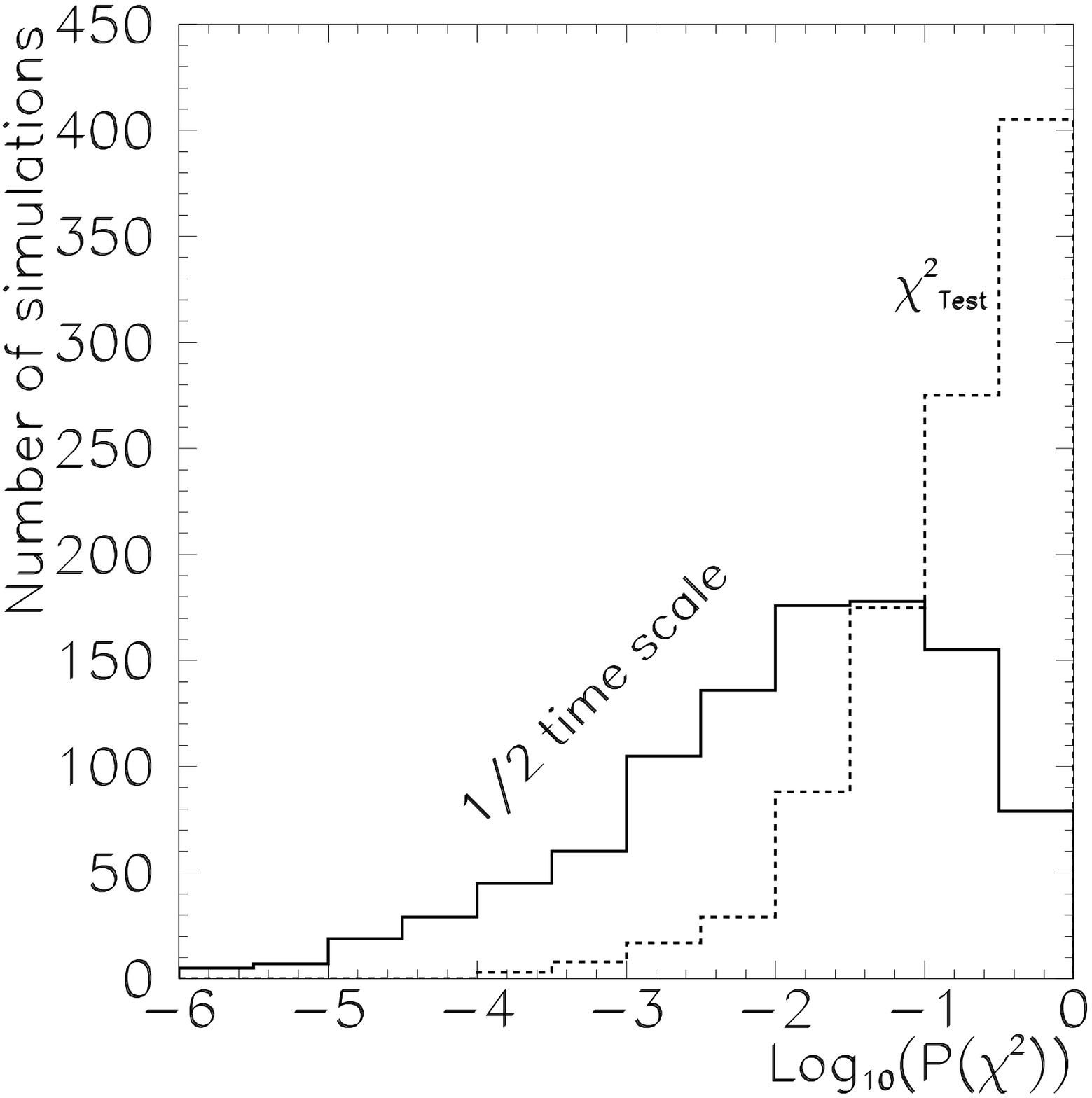}}
\rotatebox{0}{\includegraphics[scale=0.32]{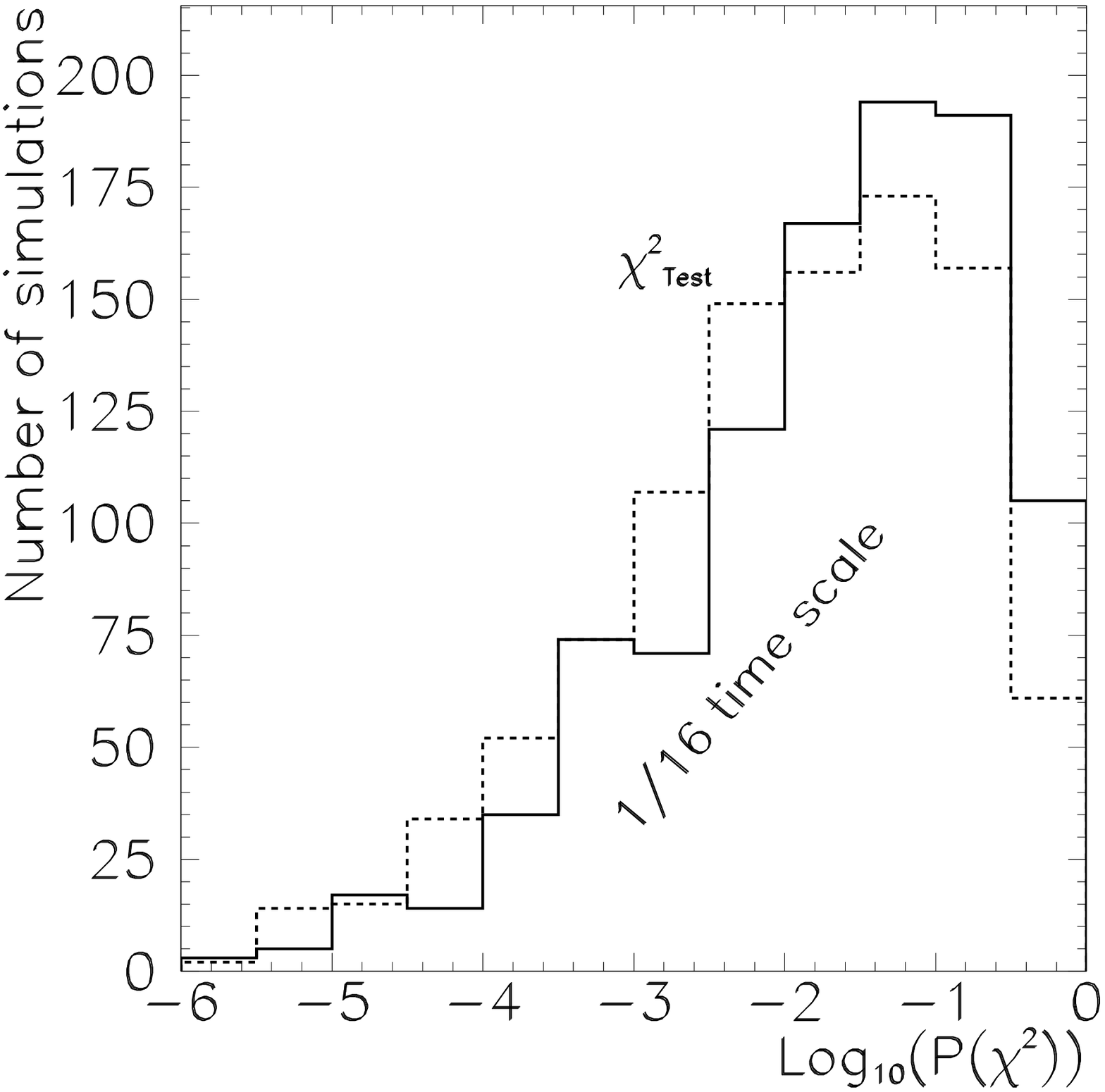}}
\end{center}
\caption{\label{chiprob} 
The distribution of $\chi^2$ probabilities are compared between the direct $\chi^2$ test for variability 
and the $\chi^2$ test with the null hypothesis for selected times scales in the Haar wavelet analysis. 
The $\chi^2$ probability distributions are shown only for the most significant timescales to avoid 
cluttering the figure. Histograms in the left panel were obtained from the simulations of a variability 
such as in Figure \ref{figure:coef1} while histograms shown on the right panel correspond to 
variability such as in Figure \ref{figure:coef4}
}
\end{figure}

In order to quantitatively characterize the differences in sensitivity noticed in the two above examples, we consider a large number (1000) of simulations of the same variability patterns and compare the distributions of the $\chi^2$ probabilities in each timescale of the Haar wavelet analysis to that of a direct $\chi^2$ test on the original data.  

The left panel of Figure \ref{chiprob} shows the distribution of the $\chi^2$ probability of the $1\over 2$ scale compared to that of a direct $\chi^2$ test in the case of a slow variability as in Figure \ref{figure:coef1}.  The $\chi^2$ probability is the complement of the CL for variability. While the direct $\chi^2$ test yields a 99\% (99.9\%) CL in 5.7\% (1.1\%) of the cases, the Haar wavelet coefficient analysis results in a 99\% (99.9\%) CL in 41.2\% (17.1\%) of the cases, demonstrating a greater sensitivity. It should be noted that when the variability is centered in the dataset instead of concentrating in one half, the variability appears in the $1\over 4$ scale with a reduced sensitivity (99\% (99.9\%) CL in 25.9\% (9.9\%) of the cases) while the $\chi^2$ test gives the same result. 

A similar comparison is shown on the right panel of Figure \ref{chiprob} in the case of the faster variability of Figure \ref{figure:coef4}. Since the variability extends over  $\sim 3$ data points, it  primarily affects the coefficients in the $1\over 16$ scale and it is the corresponding $\chi^2$ probability that is compared to the $\chi^2$ probability of the direct $\chi^2$ test for variability.  The direct $\chi^2$ test yields a 99\% (99.9\%) CL in 45.3\% (19.7\%) of the cases and the Haar wavelet coefficients analysis results in a 99\% (99.9\%) CL in 34.3\% (15.1\%) of the cases. The somewhat lower performance of the Haar wavelet coefficient analysis is another example of the sensitivity of the Haar coefficients to the position of the variability feature. When the same variability is centered on data point 38 instead of 37, both the Haar analysis and direct $\chi^2$ test result in the same $\chi^2$ and CL distributions. This suggests the CL for variability in the different timescales should be all given. 

The advantage of the Haar analysis over the direct $\chi^2$ test for longer variability timescales comes from taking into account the order of the data points. When the variability is so fast it affects only one or a very few data points, the order of the data points becomes irrelevant and both methods should have similar sensitivities as we just observed.  

\section{Application to real data}
\label{analysis}

\subsection{Complications generally associated with the analysis of real data}
     
\subsubsection{If the number of data points is not a power of two}
 The Haar wavelet analysis assumes the number of data points to be a power of two while this may not be the case with real data. A strict approach for circumventing this problem consists of truncating the dataset to a subset with a power of two as the number of data points included. This may however be frustrating when the number of data points is close but just inferior to an integer power of two. Another approach is to pad the end of the dataset with zeros until a power of two is obtained in such a way that the same algorithm as described above may be used.  The last non-zero average used in each timescale is them biased toward zero. In order to avoid this that average can be calculated counting only the data points originally in the data, excluding any padding zeros. This eliminates any data pollution caused by the zero padding. However, this also blurs the timescales as the last non zero coefficient of each timescale is calculated from a number of data points that may be different from what it is for the other coefficients in the the same timescale section.

\subsubsection{The data points have different errors}
The data points' errors were considered to be all the same in the above simulations while this is generally not the case.
In the calculation of averages, data points with the largest errors should be given a lesser weight. One can choose to use weighted averages with ${{1} \over{\delta s_i^2}}$ as the weight of data point $s_i$. The error on the average is then given by $1\over{\sqrt{\sum {1 \over {\delta s_i^2}}}}$ \cite{taylor1997}.  The wavelet coefficient errors are then obtained, just as before, as the half square root of the sum of the variances of the two averages involved. This is valid provided the  probability distributions are gaussian as we assumed from the beginning.  This was nevertheless tested by simulating 10,000 data sets of 64 samples without any variability and with the standard deviation of each data point taken randomly and uniformly between 0.1 and 10. It was then verified that the wavelet coefficients within each variability scale are normally distributed around zero with a standard deviation corresponding to the calculated coefficient error. 

In some cases, for example in counting experiments, when the counts are very finely binned in time so the individual data points derive from a Poisson statistics,  attention may have to be given to the non-gaussian nature of the measurements constituting the data. This is not investigated here and the errors are assumed to be all gaussian.

\subsubsection{Uneven data sampling}
Another difficulty arrises from the fact that data points are generally not recorded with a strict regularity.  In order to apply a Haar wavelet analysis in these situations, one could subdivide the data in powers of two according to the time data points are recorded. However, one is then likely to rapidly encounter time intervals containing no data spreading sporadically throughout the entire time interval. Alternatively, our approach consists of applying the Haar wavelet analysis as described above, ignoring irregularities in the data points' times. The resulting Haar wavelet coefficients may then not correspond to  comparison between data subsets of exactly equal durations anymore. They however still provide a useful systematic comparison between different domains of the dataset. The calculation of CL for variability as described above is still valid but the variability timescales can only be identified on the basis of a careful inspection of the data points' time distribution as illustrated in the following application example.    

\subsection{Example application}
\begin{figure}
\begin{center}
\rotatebox{0}{\includegraphics[scale=0.32]{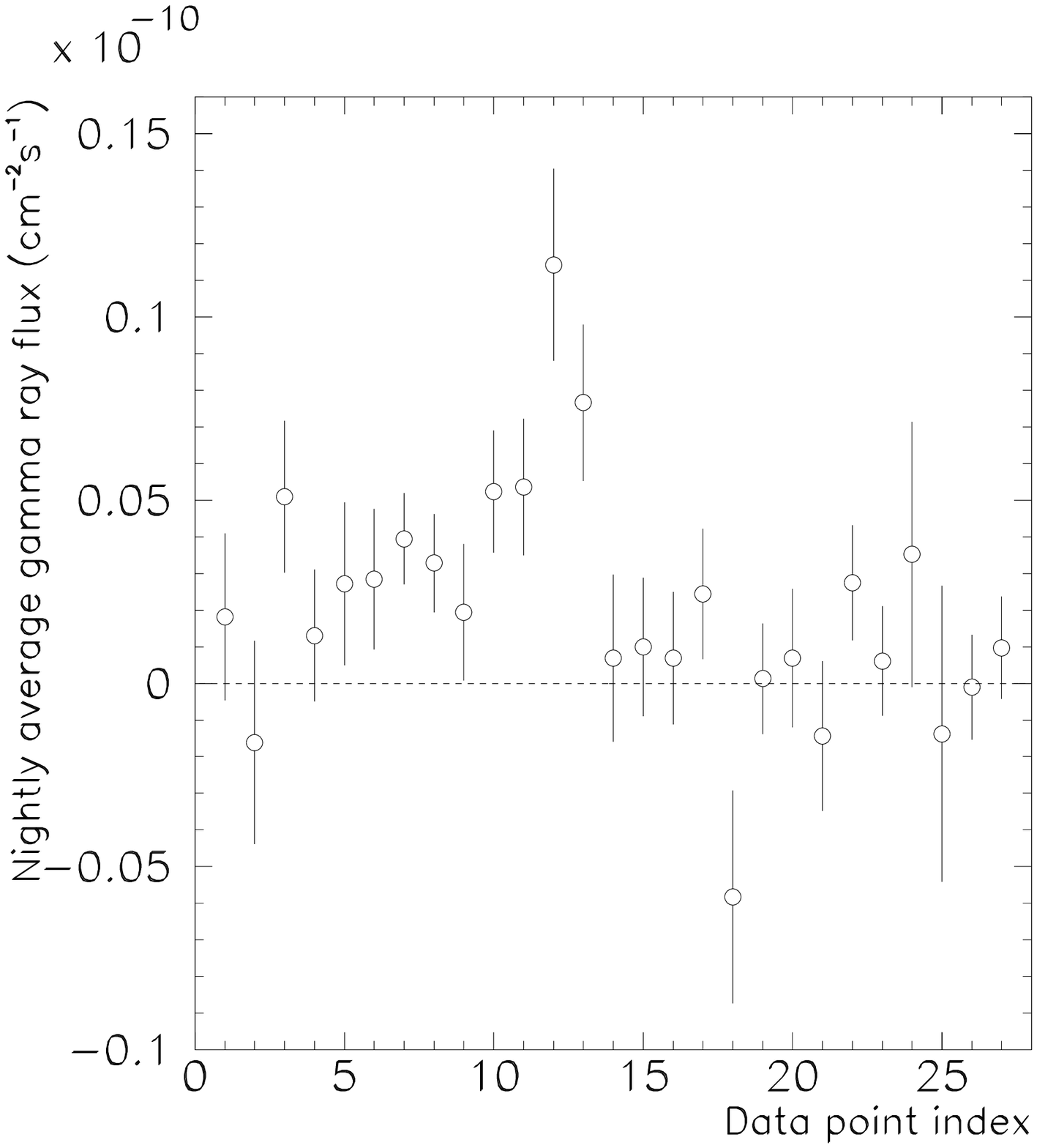}}
\rotatebox{0}{\includegraphics[scale=0.32]{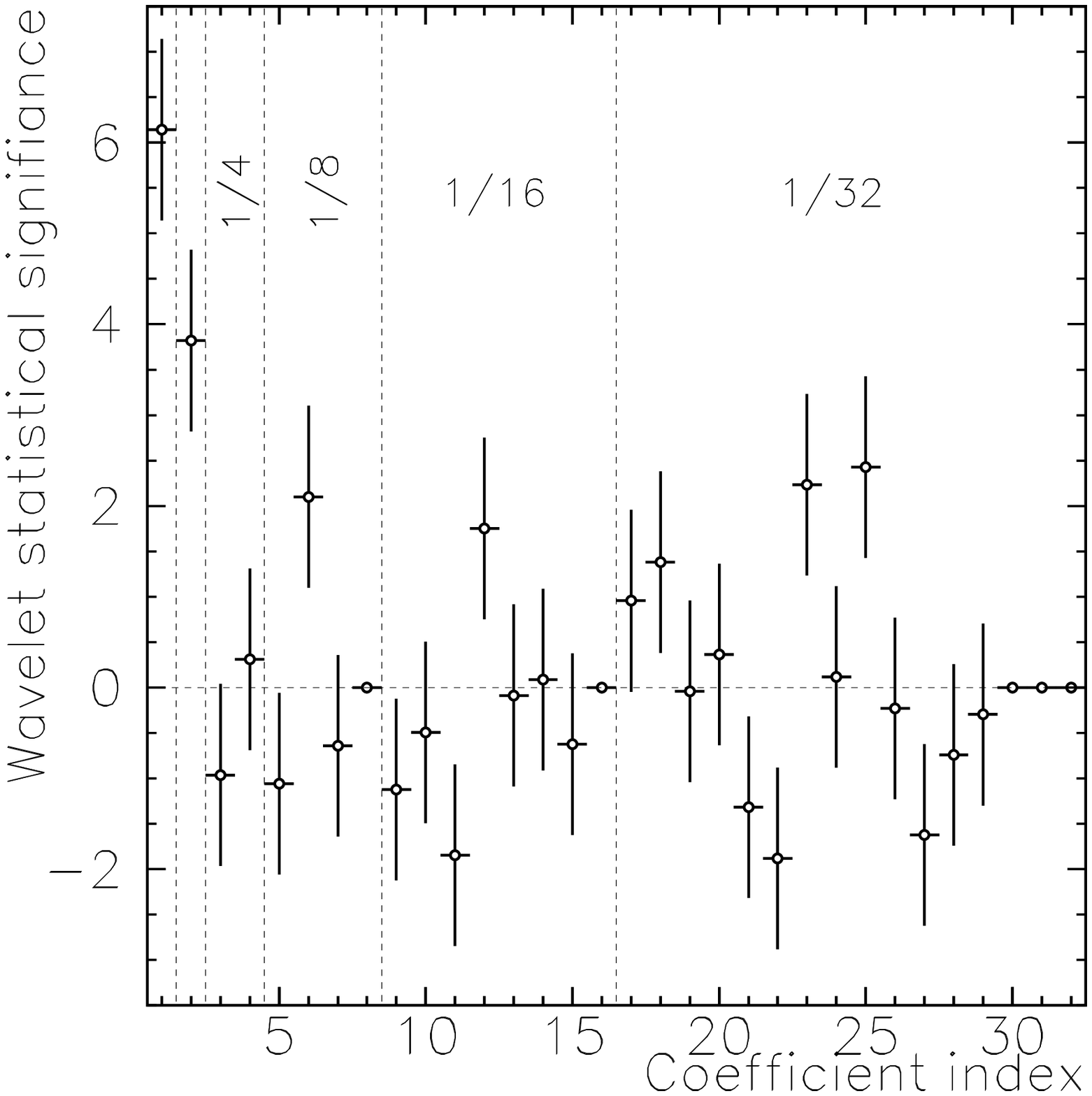}}
\end{center}
\caption{\label{m87a} Left: Light curve of M87 recorded with VERITAS at $E>250\,GeV$ with 27 measurements over the period between 12/14/2007 and 05/02/2008. It seems the average flux was higher in the first half of the dataset and that a two or three day flair may have been recorded around the $12^{th}$ measurement (02/12/2008).  Right: Statistical significance of the Haar wavelet coefficients (wavelet coefficients divided by their statistical error so all the error bars have unit amplitude) for the M87 data. See text for details.}
\end{figure}

Figure \ref{m87a} shows the light curve of M87 above $250\,GeV$ recorded with the VERITAS gamma ray observatory  with 27 nightly average measurements obtained between 12/14/2007 and 05/02/2008 \cite{m87veritas}. At first, we consider the data points in sequence, ignoring the differences between the time intervals separating them. A direct $\chi^2$ test for variability gives a 99.840\% CL indication for variability (reduced $\chi^2$ of 2.0 for 26 degrees of freedom). Without a more refined analysis,  only a visual inspection may permit to develop some idea about the variability pattern and timescales. Here, it can be seen that the data points are on average higher in the first half of the dataset and a flare may have been recorded with measurements 11, 12 and 13 which seem to be higher than their neighbors.  The CL that may be reported about the different features of the variability pattern remain undefined. 

This can in fact be clarified with the Haar coefficients analysis presented above. Since the data contains only 27 points, 5 padding zeros were added at the end. The statistical significance of the Haar coefficients is shown on the right panel of Figure  \ref{m87a}. Because of the padding with zeros, the last three coefficients of the $1\over 32$ timescale are null as well as the last of the $1\over 16$ and $1\over 8$ timescales. Table \ref{m87table} gives the CL for variability over each scale of the Haar coefficients analysis. The first coefficient, which corresponds to the average of the entire data-set,  is different from zero with a statistical significance of 6.14 which corresponds to a high CL for the detection of gamma ray emission from M87. The second coefficient, which approximately corresponds to the difference between the first and second halves of the data-set, is different from zero with a statistical significance of  3.82 corresponding to a $99.987\%$ statistical CL the flux changed.  The potential flare appears in the $1\over 16$ (difference between consecutive pairs of measurements) and $1\over 32$ (difference between consecutive measurements) scales. In the $1\over 16$  scale,  the $3^{rd}$ and $4^{th}$ coefficients present deviations from zero of just under 2 standard deviations. They correspond to differences between measurements 9 through 12 and 13 through 16 respectively. In the $1\over 32$ scale,  the $6^{th}$ and $7^{th}$ coefficients present deviations from zero of the order of 2 standard deviations. They correspond to differences between consecutive measurements 11 and 12 and measurements 13 and 14 respectively. The CLs for variability in the $1\over 16$  and $1\over 32$ timescales are found to be $70\%$ and $95\%$ respectively so the data remains inconclusive regarding the shortest timescale variability.

\begin{table}
\caption{For each timescale of the Haar coefficient analysis of the M87 data, the reduced $\chi^2$, the number of degrees of freedom (NDOF) and the corresponding CL for variability are given in the table. }
\centering
\begin{tabular}{c c c c}
\hline\hline
Scale & Reduced $\chi^2$& NDOF & CL(\%) \\ 
\hline
$1\over1$   & 37.74	& 1     & 100.0 \vspace{1mm}\\
$1\over2$   & 14.59	& 1     & 99.987 \vspace{1mm}\\
$1\over4$   & 0.51     & 2    & 40.0 \vspace{1mm}\\
$1\over8$   & 1.98	& 3     & 88.5 \vspace{1mm}\\
$1\over16$ & 1.20	& 7     & 70.2 \vspace{1mm}\\
$1\over32$ & 1.73	& 13   & 95.1 \vspace{1mm}\\
\hline
\end{tabular}
\label{m87table}
\end{table}

\begin{figure}
\begin{center}
\rotatebox{0}{\includegraphics[scale=0.32]{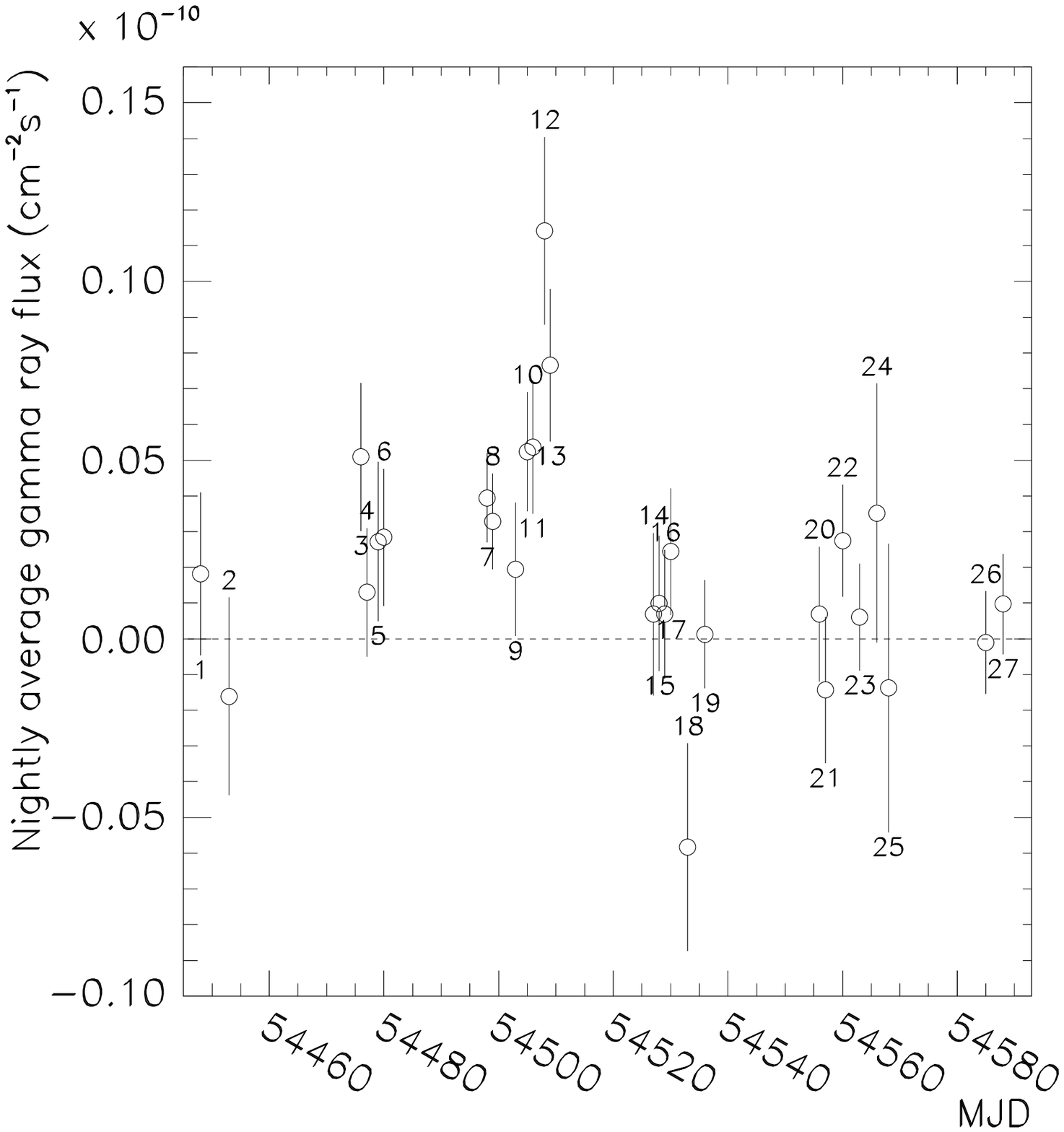}}
\rotatebox{0}{\includegraphics[scale=0.32]{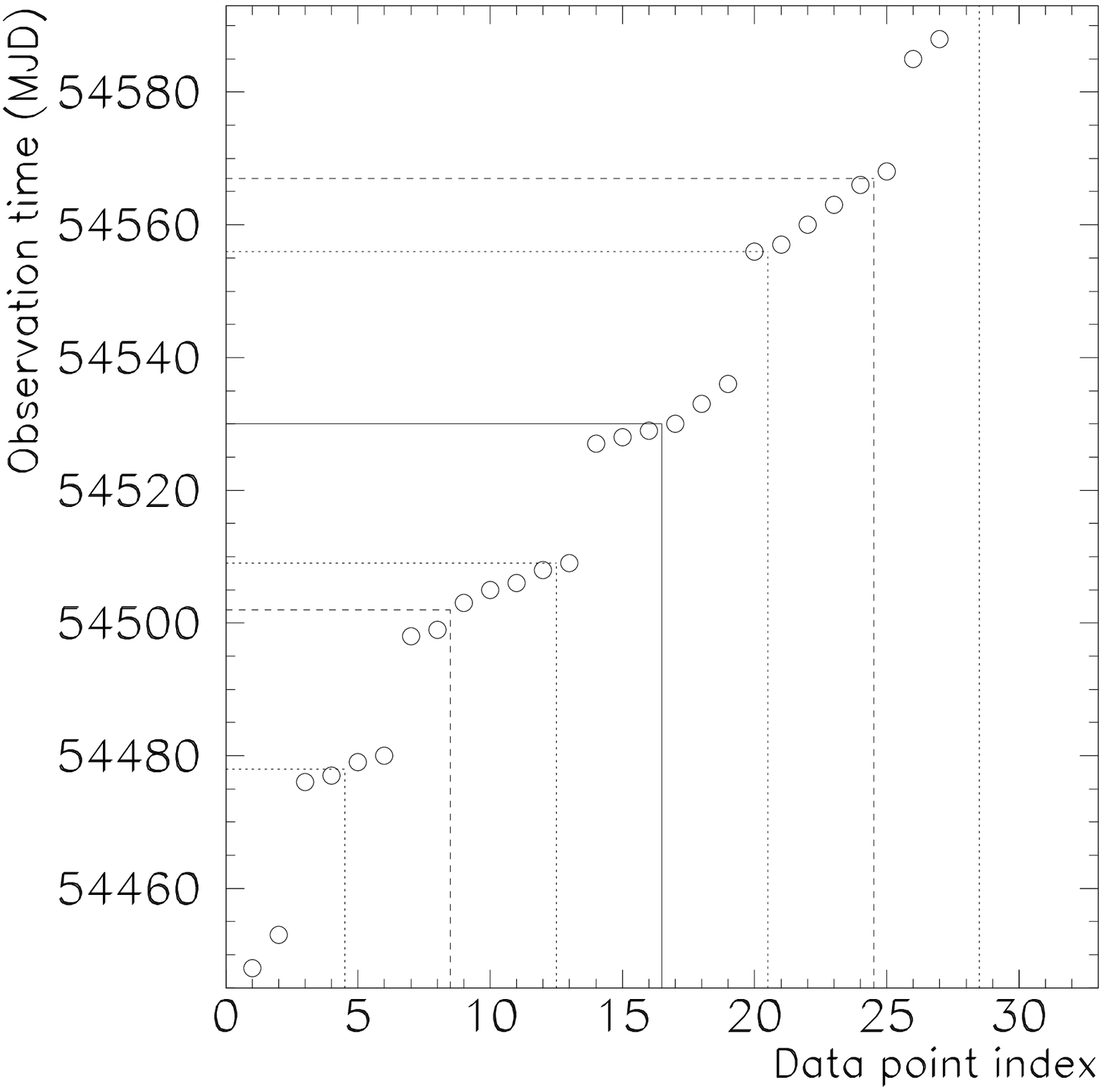}}
\end{center}
\caption{\label{m87b} Left:  M87 flux recorded with VERITAS at $E>250\,GeV$ as a function of time with the measurement index numbers indicated next to each point. Right: Measurements times presented as a function of their index. The solid, dashed and dotted line indicate the data subdivision in halves, quarters and eighths after zero padding as done in the Haar wavelet coefficients calculation. See text for details.}
\end{figure}

The data points were not recorded evenly in time. The left panel of  \ref{m87b} presents the flux measurements as a function of time and the right panel of Figure \ref{m87b} shows the modified Julian date of each observation as a function of its index. Observations are regularly interrupted by periods of full moon and, because of weather conditions,  observations were also impossible on occasional nights during moonless periods. The solid line represents the boundary between the two halves of the dataset as used in the calculation of the second Haar wavelet coefficient (the $1\over 2$ scale). The data halves have respective durations of  81 and 58 days indicating the $1\over 2$ scale corresponds to durations of $\sim 70$ days. The dashed lines subdivide the data in quarters of durations 51, 26, 36 and 20 days respectively, and the $1\over 4$ data scale is averaging to a $\sim 33$ days timescale. The dotted lines subdivide the data in seven eighths with durations of  29, 20, 5, 20, 26, 9 and 20 days respectively, and the $1\over 8$ data scale is averaging to a $\sim 18$ days timescale. It should be noted that the timescale of each wavelet domain is not strictly defined. In our example the $1\over 4$ scale spans durations from 20 to 51 days while the $1\over 8$ scale spans durations from 5 to 29 days, showing an overlap between them. Going to smaller data scales, this overlap becomes more severe and the timescales less defined. The $1\over 16$ scale spans durations from 3 to 28 days while the $1\over 32$ scale spans durations from 1 to 20 days. This makes it impossible to associate them to actual specific timescales. However, CL calculated from these data scales are still good indicators for variability over the shortest times scales available in the data. 

In our example, the short data scale CLs are too small for the few day flare to actually be claimed. However the change in flux between the first and second half of the data can be reported with a  $99.987\%$ CL, greater than the indication provided by the non-time specific $\chi^2$ test which gave a $99.840\%$ CL . Expressed in term of gaussian standard deviations, this corresponds to changing the indication for variability from less than  $3.2\sigma$ to more than $3.8\sigma$ while providing information on the variability timescale at the same time.   

\section{Conclusions}
\label{conclusion}

It is often difficult to establish a CL with which variability can be reported from the analysis of a dataset such as a light curve. A simple $\chi^2$ test may not be optimally sensitive to the long variability timescales and does not provide any indication regarding the timescales involved.  The decomposition of simulated data on the Haar wavelet basis with error propagation was shown to be a simple and useful tool for investigating variability over several times scales simultaneously. A $\chi^2$ test under null hypothesis within each timescale provides the confidence level with which variability occurs over that timescale.  This simple treatment is made possible by the fact that the Haar wavelet coefficients within one timescale are statistically independent, a property that is not shared with other wavelet families more commonly used. 

The method as described can be applied to data with irregularities, the most important of which pertain to the data points time distribution. The interpretation of the results is then complicated by the degradation of the actual definition of the timescales but the method remains usable. This was illustrated by an example application to data from the VERITAS gamma ray observatory. The sensitivity gain appears on the largest scales while at the shortest scales, both the standard $\chi^2$  test and the wavelet approach have the same sensitivity. The Haar wavelet analysis   provides a simple and systematic approach to extract variability scale information from the data.  However, it was noted that the precise identification of the variability timescales can be affected by irregularities in the time sequence of measurements. 

We have considered the case of time variability studies. The same approach could be used in the analysis of energy spectra for which the common practice consists in considering the $\chi^2$ probability of a power law fit. When the $\chi^2$ probability is large enough, there is typically no further investigation for a possible deviation from a straight  power law \cite{acciari2010c, acciari2010a}.  The Haar wavelet analysis could be applied to the residual of this power law fit to identify spectral curvature or cutoff with improved sensitivity. 

Several developments could be considered to improve this analysis. 
It was for example observed that the sensitivity of this analysis maybe affected by the precise location of the variability pattern in the data. This weakness may be alleviated by scanning the data several times, each time shifting the starting data sample. In such an analysis, additional trial factors should be taken into account in the calculation of the CLs for variability. Also, in this paper, we made use of  $\chi^2$ tests which are only justified under the assumption of the Gaussian nature of the data points errors.  A possible further development of the method could consist of replacing the $\chi^2$  with a likelihood characterization for different statistics such as Poisson. Finally, the method could easily be transposed to multi-dimensional data such as sky-maps as a tool to search for features over different scales. 

\section{Acknowledgements}
The authors are grateful to Michelle Hui and the VERITAS collaboration for providing the M87 light curve data as well as helpful suggestions. 
The authors also gratefully acknowledge support for this research under National Science Foundation Grant PHY-0947088.




\bibliographystyle{model1-num-names}
\bibliography{<your-bib-database>}

\begin{thebibliography}{00}


\bibitem{acciari2008} Acciari, V.A., et al., 2008, ApJL, 690 , 2,  L126
\bibitem{acciari2010a} Acciari, V.A., et al., 2010, ApJL, 715, L49 
\bibitem{acciari2010b} Acciari, V.A., et al., 2010, ApJL, 708 , L100
\bibitem{acciari2010c} Acciari, V.A., et al., 2010, ApJL, 709 , L163
\bibitem{m87veritas} Acciari, V.A., et al., ApJ,  716 (2010) 819-824
\bibitem{aharonian} Aharonian, F.,  et al. 2008 Rep. Prog. Phys. 71 096901
\bibitem{m87hess} Aharonian, F., et al. 2006, Science, 314, 1424
\bibitem{m87magic} Albert, J. et al. 2008, ApJ, 685, L23
\bibitem{daub} Daubechies, I., Communication on Pure and Applied Mathematics, vol. 41(1988), 909 
\bibitem{haar1910} A. Haar, Mathematische Annalen, 1910, 69, pp 331Ð371
\bibitem{optlinear}  C. H. Hsiao and W. J. Wang, Journal of Optimization Theory and Applications, 103, 3, 641-655
\bibitem{LiMa83} Li, T. P., and Ma, Y. Q. 1983, ApJ, 272, 317
\bibitem{numrec} Press, W.H., et al., Numerical recipes in C++, Second Edition, Cambridge University Press, 2002, ISBN 0-521-75033-4
\bibitem{taylor1997} Taylor, J.R., 1997, "An introduction to error analysis", University Science Books, ISBN 0-935702-75-X
\bibitem{weekes} Weekes, T.C., 2003, "Very high energy gamma-ray astronomy", IoP Publishing, ISBN 0 7503 0658 0
 \end{thebibliography}



\end{document}